# Anomalous Hall Effect in Layered Ferrimagnet MnSb$_2$Te$_4$

Gang Shi, Mingjie Zhang, Dayu Yan, Honglei Feng, Meng Yang, Youguo Shi,[*] and Yongqing Li [*]

*1. Bejing National Laboratory for Condensed Matter Physics, Institute of Physics, Chinese Academy of Science, Beijing 100190*

*2. School of Physical Sciences, University of Chinese Academy of Sciences, Beijing 100190, China*

We report on low-temperature electron transport properties of MnSb$_2$Te$_4$, a candidate of ferrimagnetic Weyl semimetal. Long-range magnetic order is manifested as a nearly square-shaped hysteresis loop in the anomalous Hall resistance, as well as sharp jumps in the magnetoresistance. At temperatures below 4 K, a ln$T$-type upturn appears in the temperature dependence of longitudinal resistance, which can be attributed to the electron-electron interaction (EEI), since the weak localization can be excluded by the temperature dependence of magnetoresistance. Although the anomalous Hall resistance exhibits a similar ln$T$-type upturn in the same temperature range, such correction is absent in the anomalous Hall conductivity. Our work demonstrates that MnSb$_2$Te$_4$ microflakes provide an ideal system to test the theory of EEI correction to the anomalous Hall effect.

---

Supported by the National Natural Science Foundation of China (Project No. 11961141011 & 61425015), the National Key Research and Development Program (Project No. 2016YFA0300600), and the Strategic Priority Research Program of Chinese Academy of Sciences (Project No. XDB28000000).

* Correspondence authors: ygshi@iphy.ac.cn, yqli@iphy.ac.cn.

The anomalous Hall effect (AHE) is one of the most fundamental transport properties of materials [1,2]. Despite having been studied for over a century, it still attracts a great deal of interest, at least partly due to recent advances in novel topological materials, such as magnetic topological insulators [3] and Weyl semimetals [4,5]. Most of the reported works on the AHE have been limited to the single particle framework, in which the many-body interaction does not play a central role [1,2]. The single particle treatment has been successful in accounting for numerous experiments that are not concerned with the transport properties at temperatures below 1 K. It was further justified by the theory of Wölfle *et al.*, who showed that electron-electron interaction (EEI) does not cause any correction to the anomalous Hall (AH) conductivity [6-8]. In a recent experiment, however, some of us found that the EEI may underlie pronounced low temperature corrections to the AH conductivity in magnetic semiconductor $HgCr_2Se_4$ [9]. In contrast, earlier experiments on ferromagnetic metals or semiconductors found either no quantum correction to the AH conductivity [10,11], or finite corrections that were attributed to the weak localization [8,12-14]. In a latest theoretical work, Li and Levchenko showed that the EEI correction to the AH conductivity could exist [15], but the predicted temperature dependence is inconsistent with the experiment reported in Ref. [9]. To resolve these discrepancies, it is necessary to expand the low temperature AHE measurement to more materials with broken time-reversal symmetry.

In this work, we perform electron transport measurements of $MnSb_2Te_4$ thin flakes exfoliated from the bulk single crystals. At low temperatures, both the ordinary and the anomalous Hall components have been observed, and these two components can be separated straightforwardly due to nearly square-shaped hysteresis curves and low coercivity fields. At temperatures under 5 K, both the longitudinal and AH resistances deviate from the Bloch-law-like behavior observed at higher temperatures, and can be attributed to the EEI correction. However, the AH conductivity is found to follow the Bloch-law-like behavior down to at least 0.4 K. These results will be explained with the theory of EEI correction to the AHE.

MnSb$_2$Te$_4$ belongs to a family of ternary telluride AB$_2$Te$_4$, which has a layered structure shown in Fig. 1a. Its close cousin, MnBi$_2$Te$_4$, is an antiferromagnetic topological insulator [16-19], and has recently attracted a lot of attention because of the realization of the quantized AHE and possibly other topological states in the exfoliated thin flakes [20,21]. At low temperatures, each septuple layer (Te-Bi-Te-Mn-Te-Bi-Te) of MnBi$_2$Te$_4$ has a ferromagnetic order with out-of-plane anisotropy, and couples antiferromagnetically to adjacent quintuple layers via superexchange interaction [17,22]. MnSb$_2$Te$_4$, in contrast, has a ferrimagnetic order with a Curie temperature of 25 K, according to a recent work reported by Murakami *et al.* [23]. They proposed that this material may be a type II Weyl semimetal, and the presence of Mn-Sb intermixing leads to ferromagnetic coupling between the majority-spin Mn layers, with help of the minority Mn spins on the Sb-substitutional sites [23].

The MnSb$_2$Te$_4$ single crystals used in this work were synthesized with the flux method. Starting materials Mn (piece, 99.99%), Sb (grain, 99.9999%) and Te (lump, 99.9999%) were mixed in an Ar-filled glove box at a molar ratio of 1 : 10 : 16. The mixture, placed in an alumina crucible, was sealed in an evacuated quartz tube. The tube was heated to 700 °C for over 10 h and dwelt for 20 h in a furnace. It was then cooled down to 630 °C at a rate of 0.5 °C/h, followed by centrifuge separation. The obtained single crystals were exfoliated to microflakes of a few tens of nanometers in thickness by using the Scotch tape method. After electron beam lithography, ohmic contacts were formed by e-beam deposition of Pd/Au thin films and a liftoff process. Fig. 1(b) shows the optical image of one of the five samples studied in this work. Electron transport measurements were performed in a $^3$He cryostat or a $^4$He cryostat by using standard lock-in technique with electrical wirings schematically depicted in Fig. 1(c). The magnitude of source-drain current was mostly set at 100 nA to avoid heating or other spurious effects. The data shown below were taken from an 18-nm-thick MnSb$_2$Te$_4$ flake, and the similar results were also obtained from other samples. Symmetrization and anti-symmetrization procedures were applied to the raw longitudinal and Hall resistances to remove the mixing of the two transport components, which arises primarily from misalignment of Hall probes, as well as the irregular sample shape.

Fig. 2(a) shows the temperature dependence of longitudinal resistance $R_{xx}$ [24]. A resistance minimum appears at 5 K, above which $R_{xx}$ has a linear dependence on $T^{5/2}$, as depicted in Fig. 2(b). Similar $T^{5/2}$ dependence was also observed previously in other magnetic systems, such as $La_{1-x}Ca_xMnO_3$ [25] and $CrO_2$ [26]. This was attributed to a combination of electron-electron, electron-magnon and electron-phonon scatterings, which can lead to a power-law temperature dependence $T^\alpha$ with exponent $\alpha$ in a range of 2-5 [25]. Subtracting the $T^{5/2}$ background from the measured resistance yields the $\Delta R_{xx}$ - $T$ plot displayed in Fig. 2(c). It shows that the low temperature upturn at $T < 4$ K has a logarithmic temperature dependence. For a thin film in the diffusive transport regime, either magnetic or non-magnetic, ln$T$-type correction to $R_{xx}$ has often been observed. It can be explained satisfactorily with EEI, weak localization or a combination of the two mechanisms [27,28]. As delineated below, the weak localization effect can be excluded in our $MnSb_2Te_4$ samples, based on the magnetoresistance (MR) measurements.

Fig. 3(a) shows the magnetic field dependence of $R_{xx}$ between $B = \mu_0 H = +9$ T and -9 T, with both up and down field-sweep directions. The main feature is the W-shaped magnetoresistance with minima located at $\mu_0 H = \pm 4.2$ T, in addition to the sharp jumps at $\mu_0 H = \pm 0.12$ T. In Fig. 3(b), the sudden drops in $R_{xx}$ can be seen more clearly. They can be attributed to a magnetization reversal process, which takes place in a narrow range of magnetic fields ($\Delta B \sim 0.02$ T). Fig. 3(c) shows the magnetoresistances, defined as MR=($R_{xx}$(B)/$R_{xx}$(0)−1)×100%, at several temperatures. Only the data for the magnetic fields swept from above magnetization saturation toward zero are included, so that these MR curves are not influenced by magnetization reversal. At the first sight, the cusp-shaped negative MR at low fields is similar to that expected for the weak localization [27,28]. The magnitude of MR, however, varies very little at temperatures below 2 K, and increases from 1% to 3% as $T$ is varied from 2 K to 12 K. Such a temperature dependence is *opposite* to that of the weak localization effect, in which the resistance correction arises from quantum coherence and would become weaker at higher temperatures. Therefore, the weak localization is unlikely responsible for the negative MR in MnSb$_2$Te$_4$. Instead, it can be attributed to magnetic field suppression of spin disorder, which is commonly seen in magnetic systems. It is also noteworthy that the negative MR can persist to the magnetic fields well beyond the magnetization hysteresis loop. This suggests that the spin disorder underlying the negative MR cannot be ascribed to the domain wall dynamics. This is probably associated with atomic scale defects, such as the randomly distributed Mn spins on the Sb-substitution sites, which may not align perfectly antiparallel to the majority Mn-spins located in the center plane of the MnSb$_2$Te$_4$ septuple layer because of the local disorder.

Fig. 4(a) displays the Hall resistance data recorded at $T = 2$ K, which contains both the ordinary and the anomalous Hall components. The former is nearly linear, with a Hall coefficient of about 1 $\Omega$/T, corresponding to a hole density of ~3.5×10$^{20}$ cm$^{-3}$, from which the carrier mobility is estimated to about 10 cm$^{-2}$/V·s. The slight curvature in the ordinary Hall effect may be attributed to coexistence of a small number of electron-type carriers with the holes, consistent with the semimetallic band structure reported in Ref. [23]. The dominance of the hole-type carriers may be related to intermixing of Sb and Mn sites, which lowers the chemical potential, and makes electron pocket much smaller than that of a defect-free sample. Nevertheless, the Mn-Sb intermixing is essential for the AHE studied in this work. According to Ref. [23], about 1/3 of Mn spins are on the Sb-substitutional sites and they align antiparallel to the Mn spins on the original sites (in the middle of the septuple layer, see Fig. 1(a)). Antiferromagnetic coupling between these two types of Mn spins is thus believed to be responsible for the proposed ferrimagnetic ordering. It is noteworthy, however, that the carrier density in $MnSb_2Te_4$ is on the order of $10^{20}$ cm$^{-3}$, which is comparable to many diluted magnetic semiconductors [29], so it is still unclear whether the magnetic order is influenced by carrier-mediated interactions.

Fig. 4(b) shows that there exists a nearly square-shaped hysteresis loop in each AH resistance curve at low temperatures. The extracted coercive fields are same as those obtained from the sharp jumps in the MR curves. A striking feature is that the $R_{xx}$ maxima do not appear at the coercive fields, and they are rather located at the onset of magnetization reversal. This strongly suggests that the magnetization dynamics in $MnSb_2Te_4$ is different from those encountered in common ferromagnetic metals, in which the resistance maxima are related to magnetic domains or domain walls. Fig. 4(c) also shows that both the coercive field and the AH resistance decrease rapidly with increasing temperature. A linear extrapolation of the coercive fields at $T$ = 10 -30 K yields $T_C$ = 36.5 K. It is considerably higher than the 25 K obtained from the magnetization measurements of polycrystalline $MnSb_2Te_4$ samples in Ref. [23]. The Curie temperature extracted from our AHE measurements is, however, consistent with the magnetization measurements of the single crystal samples from the same batch. The higher $T_C$ in our samples may be attributed to a different level of Mn-Sb intermixing or less defects.

Fig. 4(d) displays temperature dependence of the AH resistance at $B = 0.2\,\text{T}$. This clearly deviates from the usual Bloch-law-like behavior at $T < 4$ K, and the deviation becomes more pronounced as the temperature approaches to zero. This is in contrast to the temperature dependence of the AH conductivity, which is converted from the AH resistance with $\sigma_{AH} = R_{AH}/[R_{xx}^2 + R_{AH}^2]$. As depicted in Fig. 4(e), the temperature dependence of $\sigma_{AH}$ is similar to those commonly seen in ferromagnetic materials. As depicted in Fig. 4(f), the low field AH conductivities are weakly dependent on the temperature for $T < 4$ K. Fig. 4(g) further shows that the AH conductivity at $B = 0.2$ T follows $\sigma_{AH} \propto 1 - (T/T_C)^\alpha$ with $\alpha = 2.02 \pm 0.02$ for a wide range of temperatures (*T*=0.4-33 K). Here the $\alpha$ value is greater than α=3/2 in the Bloch law. The larger $\alpha$ value, is, however, not rare in literature. For instance, in Ref. [30], $\alpha \approx 2$ was obtained for (La,Sr)MnO$_3$ and (La,Ca)MnO$_3$ samples. Linear extrapolation to zero $\sigma_{AH}$ in Fig. 4(g) produces a transition temperature of 34.9 K, which is only 4 % smaller than the $T_C$ value obtained from the *T*-dependence of coercive field (Fig. 4(c)).

The validity of $\sigma_{AH} \propto 1 - (T/T_C)^{2.02}$ down to the lowest temperature (0.4 K) suggests that there is no low temperature correction to $\sigma_{AH}$ due to either EEI or weak localization in MnSb$_2$Te$_4$, since both effects would be more pronounced at lower temperatures. The absence of quantum correction to $\sigma_{AH}$ has also been observed before in ferromagnetic metals and semiconductors (e.g. Fe thin films [10], (Ga,Mn)As thin films [11], and theoretically explained by considering the skew scattering and side jump mechanisms, as well as a symmetry requirement [6,7]. In Refs. [10] and [11], however, no measurement has been reported to exclude the weak localization effect, which can also modify the AH conductivity [7,31].

The absence of the low-temperature correction to $\sigma_{AH}$ can give a straightforward explanation of the low temperature upturn in $R_{AH}$ at $T < 4$ K shown in Fig. 4(d). Since $\sigma_{AH} \approx \frac{R_{AH}}{R_{xx}^2}$ [32], the relative changes in $R_{AH}$ and $R_{xx}$, must satisfy $\delta R_{AH} = \frac{\Delta R_{AH}(T)}{R_{AH,0}} \approx 2\delta R_{xx} = 2\frac{\Delta R_{xx}(T)}{R_{xx,0}}$ in order to yield $\Delta \sigma_{AH} \approx 0$. Here $\Delta R_{AH}(T) = R_{AH}(T) - R_{AH,0}(T)$, $\Delta R_{xx}(T) = R_{xx}(T) - R_{xx,0}(T)$, and $R_{AH,0}$ ($R_{xx,0}$) refers to the $R_{AH}$ ($R_{xx}$) value without the EEI correction. As shown in Fig. 4(h), $\delta R_{AH}$ is indeed twice as large as $\delta R_{xx}$, and both of them have a $\ln T$-type temperature dependence. Therefore, the low temperature corrections to the AHE is fully in agreement with the theory reported by Wölfle *et al.* in Refs. [6,7].

In summary, we have investigated the transport properties of MnSb$_2$Te$_4$ thin flakes, which provide an ideal platform to study quantum corrections to AHE, because of their low magnetization saturation fields and nearly square-shaped hysteresis loop. Here $\ln T$-type corrections are observed for both longitudinal and AH resistivities at $T < 4$ K, quantum correction to the AH conductivity is negligible. Since the weak localization effect can be ruled out by the temperature dependence of MR, the low temperature corrections to longitudinal and AH resistivities can be attributed to the EEI effect first identified by Altshuer and Aronov [33]. In particular, the absence of quantum correction to the AH conductivity is in excellent agreement with the EEI theory subsequently adapted for the AHE [6,7]. Experimental efforts on more materials are highly desirable in order to observe the anomalously large quantum corrections to both AH resistivity and AH conductivity found in $HgCr_2Se_4$ [9]. This may help to gain a deeper insight into the AHE in the framework of many body physics.

**Acknowledgments:**

We thank technical support of Rencong Zhang, Meng Yang and Tianzhong Yang, and fruitful discussions with Hongyi Xie .

[32] This approximation is valid because $R_{xx}$ is two orders of magnitude larger than $R_{AH}$ in MnSb$_2$Te$_4$. In this work, the AH conductivity $\sigma_{AH}$ refers to the sheet conductance per square, and hence has a dimension of inverse resistance. It is convenient for discussing the physics at two dimensions.

**Figures and captions**

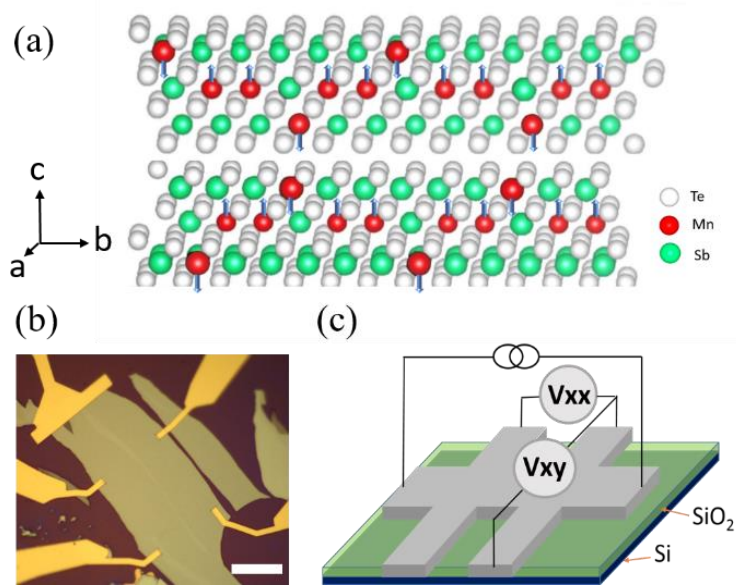

Fig. 1 Crystal structure of MnSb$_2$Te$_4$ and sample geometry. (a) Crystal structure of MnSb$_2$Te$_4$. (b) Optical image of a typical device based on a thin flake exfoliated from a bulk single crystal. The scale bar is 20 μm. (c) Schematic sketch of the device used for recording the electron transport data shown in Figs. 2-4.

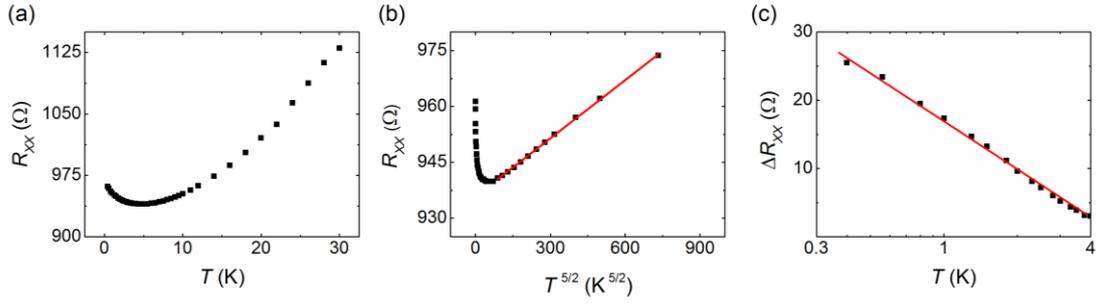

Fig. 2 Temperature dependence of longitudinal resistance $R_{xx}$ in the MnSb$_2$Te$_4$ microflake. (a) $R_{xx}$ plotted as a function of $T$. (b) $R_{xx}$ as a function of $T^{5/2}$ (symbols) and a linear fit for $T > 6$ K (line). (c) $\Delta R_{xx}$ vs T, where $\Delta R_{xx}$ is the difference between the raw $R_{xx}$ value and the linear high temperature background shown in panel (b).

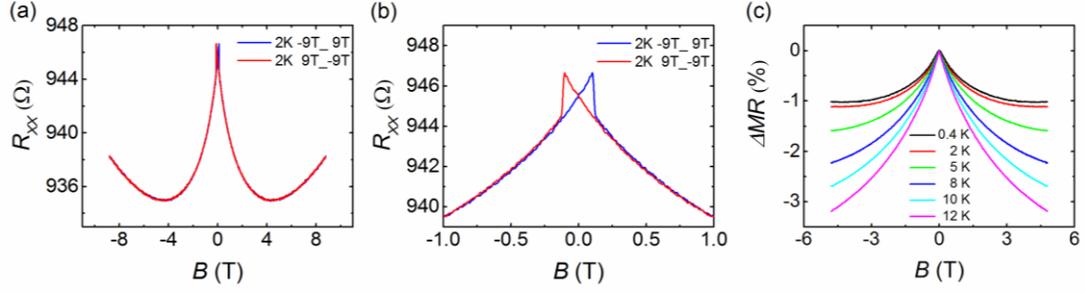

Fig. 3 Magnetoresistance in the MnSb$_2$Te$_4$ microflake. (a) $R_{xx}$ as a function of magnetic field for $\mu_0 H$ between −9 and +9 T. (b) $R_{xx}$ for $|B| < 1$ T. (c) Magnetoresistance for $|B| < 5$ T at $T$ = 0.4-12 K, in which only the sweeps from high magnetic fields toward zero field are included to avoid the complications from magnetization reversal. The data in panels (a) and (b) were taken at $T$ = 2 K.

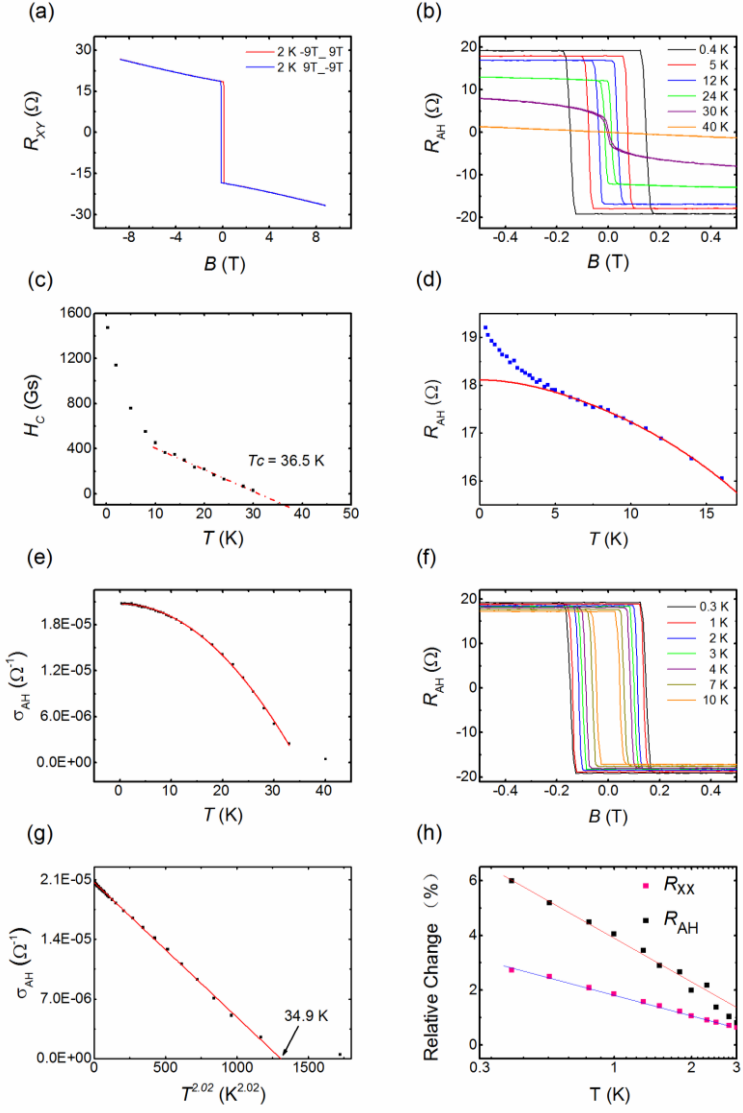

Fig. 4 Anomalous Hall effect in the MnSb$_2$Te$_4$ microflake. (a) Hall resistance as a function of magnetic field at $T$= 2 K. (b) Hysteresis loops in the AH resistances at temperatures from 0.4 to 40 K. (c) Temperature dependence of coercive field $H_c$. (d) $T$-dependence of $R_{AH}$ (symbols) and the Bloch-law-like background obtained from a fit to the data at $T$ = 6-16 K (line). (e) Behavior of $\sigma_{AH}$ as a function of $T$. (f) Hysteresis loops in the AH resistances at temperatures below 10 K. (g) Characteristics of $\sigma_{AH}$ as a function of $T^\beta$, with $\beta = 2.02$. (h) Relative deviation of the AH resistance from the Bloch-law-like behavior, $\delta R_{AH}$, plotted as a function of $T$, along with the $R_{xx}$ counterpart (see text for details).